\documentclass[aoas,MSNbibl,nameyear,dvips]{arximspdf}
\usepackage{dcolumn}
\usepackage{graphicx}

\doi{10.1214/11-AOAS519}
\volume{6}
\issue{2}
\pubyear{2012}
\firstpage{625}
\lastpage{651}

\makeatletter
\newcolumntype{d}[1]{D{.}{.}{#1}}
\newcommand{\eqref}[1]{(\ref{#1})}
\renewcommand{\citep}[1]{\citeauthor{#1} \citeyear{#1}}
\makeatother

\begin{document}
\begin{frontmatter}

\title{Investigating international new product diffusion speed: A
semiparametric approach}
\runtitle{International new product diffusion speed}

\begin{aug}
\author[A]{\fnms{Brian M.} \snm{Hartman}\corref{}\thanksref{t1}\ead[label=e1]{brian.hartman@uconn.edu}\ead[label=u1,url]{http://homepages.uconn.edu/\textasciitilde brh11004}},
\author[B]{\fnms{Bani K.} \snm{Mallick}\thanksref{t2}\ead[label=e2]{bmallick@stat.tamu.edu}}
\and
\author[C]{\fnms{Debabrata}~\snm{Talukdar}\thanksref{t3}\ead[label=e3]{dtalukda@buffalo.edu}}
\thankstext{t1}{Supported by National Science foundation CMG research
Grants DMS-07-24704.}
\thankstext{t2}{Supported in part by National Science foundation CMG
research Grants DMS-07-24704, DMS-09-14951 and by Award Number KUS-CI-016-04
made by King Abdullah University of Science and Technology (KAUST).}
\thankstext{t3}{Supported in part through the Dean's Faculty Research
Fellowship award from the School of Management, State University of New
York at Buffalo.}
\runauthor{B. M. Hartman, B. K. Mallick and D. Talukdar}
\affiliation{University of Connecticut, Texas A\&M University and State~University~of~New~York at Buffalo}
\address[A]{B. M. Hartman\\
Department of Mathematics\\
University of Connecticut\\
196 Auditorium Road\\
Unit 3009\\
Storrs, Connecticut 06269-3009\\
USA\\
\printead{e1}\\
\printead{u1}}
\address[B]{B. K. Mallick\\
Department of Statistics\\
Texas A\&M University\\
3143 TAMU\\
College Station, Texas 77843-3143\\
USA\\
\printead{e2}} 
\address[C]{D. Talukdar\\
School of Management\\
State University of New York at Buffalo\\
215E Jacobs Management Center\\
Buffalo, New York 14260-4000\\
USA\\
\printead{e3}}
\end{aug}

\received{\smonth{4} \syear{2010}}
\revised{\smonth{10} \syear{2011}}

%
\begin{abstract}
Global marketing managers are interested in understanding the speed of
the new product diffusion process and how the speed has changed in our
ever more technologically advanced and global marketplace.
Understanding the process allows firms to forecast the expected rate of
return on their new products and develop effective marketing
strategies. The most recent major study on this topic
[\textit{Marketing Science} \textbf{21} (2002) 97--114] investigated
new product diffusions in the United
States. We expand upon that study in three important ways. (1) Van den
Bulte notes that a similar study is needed in the international
context, especially in developing countries. Our study covers four new
product diffusions across 31 developed and developing nations from
1980--2004. Our sample accounts for about 80{\%} of the global economic
output and 60{\%} of the global population, allowing us to examine more
general phenomena. (2) His model contains the implicit assumption that
the diffusion speed parameter is constant throughout the diffusion life
cycle of a product. Recognizing the likely effects on the speed
parameter of recent changes in the marketplace, we model the parameter
as a semiparametric function, allowing it the flexibility to change
over time. (3) We perform a variable selection to determine that the
number of internet users and the consumer price index are strongly
associated with the speed of diffusion.
\end{abstract}

%

\end{frontmatter}
\newpage

\section{Introduction}\label{sec1}
The diffusion process of a new product describes the growth in the
product's penetration level, the proportion of the relevant population
who has adopted the new product [\citet{Bass1969}]. For global business
managers, a key issue of interest has always been the diffusion process
of new products with in and across countries
[\citet{Chandrasekaran2007};
\citet{Talukdar2002}]. The recent unprecedented globalization of the
marketplace has only heightened that interest. According to the World
Bank (\citeyear{Worldbank2010}), the volume of trade and direct investments internationally
grew by about 126\% and 550\%, respectively, from 1990--2007. As
businesses pursue new international market opportunities in an
increasingly ``flat world'' [\citet{Friedman2005}], an especially
interesting aspect of international marketing is the speed of new
product diffusions [\citet{Kohli1999}; \citet{Peres2010}; \citet{VandenBulte2000}]. Is
there any systematic trend in the speed of the international diffusion
of new products over the recent decades? Which factors hasten or slow
the process? Insights to these questions hold significant implications
for strategic planning of investments for development and introduction
of new products [\citet{Putsis1997}; \citet{Talukdar2002}].

Not surprisingly, given its strategic importance to businesses, there
has been a steady stream of studies in new product diffusion [for a
good review of this literature, refer to \citet{Chandrasekaran2007} and
\citet{Peres2010}]. This stream of studies primarily focuses on
developing and empirically testing predictive models. Typically, these
studies use country-specific but time-invariant covariates for
diffusion speed parameters to analyze spatial or across-country
variation [\citet{Talukdar2002}]. However, when it comes to the specific
issue of investigating systematic change over time in the speed of new
product diffusion, the literature is quite limited [\citet{Peres2010}].
\citet{VandenBulte2000} provides a nice review and critique of this
limited stream of literature.

As \citet{VandenBulte2000} notes, the existing insights on the issue of
diffusion speed change are often based on anecdotal evidence from the
business press rather than systematic studies. He further points out
that the few academic studies in this area typically suffer from
shortcomings in their analysis and from the limited scope of their
data. For instance, these studies use no formal or use statistically
weak methodologies to test for diffusion speed change over time [e.g.,
\citet{Fisher1971}; \citet{Grubler1990}; \citet{Clark1984}]. Also, they mainly use data
from before the public introduction of the internet and in the United
States only. Within this limited set of existing studies, the study by
\citet{VandenBulte2000} represents the most rigorous investigation of
new product diffusion speed change to date. Our study extends that
study in several important ways---both substantively and methodologically.

First, the scope and generality of the findings from the study by \citet
{VandenBulte2000} is limited by the fact that its data only includes
new product diffusions within the United States and only through 1996,
before the popular emergence of the internet. As Van den Bulte himself
notes, an important research need is a similar study in an
international context, especially in developing countries. Recent
reviews of the new product diffusion literature also underscore the
need for studies that expand the scope to include developing countries
[\citet{Chandrasekaran2007}; \citet{Peres2010}]. Our study works to fill that
need. We cover four new product diffusions in each of 31 developed and
developing nations from 1980--2004. Our set of 31 countries accounts for
about 80\% of the global economic output and 60\% of the global
population. The time period of our analysis also encompasses several
interesting and relevant world events---for example, the global
economic slow-down and stock-market crash from the 1980s, the end of
the cold war, and the popular emergence of the internet in the
mid-1990s---in the context of investigating change in new product
diffusion speed over time.

Second, our study not only provides the needed counterpart in terms of
global and post-internet era scope to the study by \citet
{VandenBulte2000}, but also uses novel methodological approaches to
analyze changes in diffusion speed. Specifically, the model used in
\citet{VandenBulte2000} makes the restrictive assumption that
consumers' propensity to adopt a new product remains constant over its
diffusion life cycle. In the logistic diffusion model, that is,
equivalent to constraining the diffusion speed parameter to be time
invariant [\citet{Dixon1980}]. While this assumption makes the empirical
estimation of the model parameters simpler, it comes at the cost of
imposing the unrealistic premise that consumers would necessarily
exhibit the same propensity to adopt a new product in the early phases
as in the later phases of its diffusion life cycle. In contrast, we
adopt a semiparametric model structure that allows the diffusion speed
parameter to vary over the diffusion life cycle of a new product.

It is relevant to point out here that there are previous studies [e.g.,
\citet{VanEverdingen2005}; \citet{Xie1997}] in the new product diffusion
literature that have also allowed time-varying diffusion speed
parameters. However, such studies are very limited in number [\citet
{VanEverdingen2005}]. More importantly, the focus of this limited set
of studies is primarily on methods which allow for time-varying
diffusion parameters to reduce the out-of-sample prediction error. The
studies show that allowing for time-varying parameters does indeed help
their models to improve the prediction of future adoptions. However,
none of them discuss the parameters' temporal patterns, that is, how
the parameters themselves changed over a product's diffusion cycle,
other than the difference between their initial (before any data)
estimates and the final estimates. Therefore, we cannot specifically
compare our findings on parameters' temporal patterns to those from the
aforesaid studies. Further, the data used by those studies is quite
limited in its scope. For instance, \citet{Xie1997} use data from the
pre-internet time period and only within the United States. Similarly,
\citet{VanEverdingen2005} use data from the very early phases of the
internet era and only within a small and similar group of developed
countries in Europe.

Finally, our study also uses a variable selection procedure to develop
a~parsimonious model from the multitude of potential country-specific
covariates available in an international diffusion study. Such
data-driven selection of a parsimonious set of country-specific
covariates is particularly valuable to business managers when deciding
which relevant market indicators to track in a global marketplace.

Taken together, the scope of our data and our methodology enable us to
shed insights into several important time-relevant issues that are
hitherto missing from the literature on new product diffusions. They
include the following: What systematic patterns do we see in terms of
change in international new product diffusion speed since 1980? What
are the macro-environmental factors related to global new product
diffusion speed patterns? To what extent are such patterns due to
changes in the levels of certain country-specific macro-environmental
factors versus a change independent of those factors? As the global
marketplace has experienced major socio-economic and technological
changes over the past three decades with likely consequences on
consumers' propensity to adopt new products, insights into the
aforesaid questions are especially interesting to both researchers and
business managers.

The next section describes our data used in this study. Section \ref{secMethod} and
Section \ref{sec4} provide details of our estimation methodology. Section~\ref{sec5}
explains the results and the \hyperref[app]{Appendix} concludes.

\section{Data}\label{sec2}
As noted earlier, the new product diffusion data used in our study
consists of four product categories across 31 countries. The product
categories are CD players, camcorders, home computers and cellular
phones. Data collection for international new product diffusion studies
has always been a challenging task [\citet{Chandrasekaran2007}]; our own
experience in the context of this study proves no exception. The key
data for analyzing international new product diffusion is the annual
product penetration level---that is, the proportion of the relevant
population which has adopted a new product. Ideally, researchers would
like to collect the annual product penetration data directly. However,
often such data is not directly available, especially for developing
countries [\citet{Talukdar2002}]. In such cases, researchers use the
more readily available annual product\vadjust{\goodbreak} sales data to indirectly compute
the corresponding annual product penetration levels as the ratio of the
product sales to population levels. However, when using indirectly
computed product penetration levels from sales data, it is important to
mitigate any potential contamination due to the inclusion of
replacement purchases as opposed to only adoption or first purchases in
product sales data [\citet{VandenBulte2000}].

Accordingly, like the existing international diffusion studies [\citet
{Putsis1997}; \citet{Talukdar2002}], we use direct annual penetration level
data whenever it is available, and use indirect or computed penetration
level data otherwise. In our set of four product categories, we were
able to get direct penetration data for cellular phones and home
computers, but had to use sales data to estimate the penetration for
camcorders and CD players. At the same time, as has been the practice
in the existing diffusion studies [\citet{Talukdar2002}], we use sales
data only from within the first seven years of respective product
introductions in a country for camcorders and CD players to reduce the
contamination of replacement purchases on our estimates. As such, while
we have an average of 17 years of data per country for cellular phones
and homes computers, we only have 7 years data per country for
camcorders and CD players. The overall time period covered by our
diffusion data for the four product categories across the selected 31
countries spans 25 years from 1980 to 2004. For the individual product
categories, the time periods covered are as follows: CD players
(1985--1993), camcorders (1987--1996), home computers (1980--2004), and
cellular phones (1980--2002).

Table \ref{tablecountries} below lists the 31 countries that we use in
our study. As the list shows, it consists of most of the major
developed and developing countries and accounts for about 80\% of the
world economic output and 60\% of the world population. Thus, our study
has 124 (4${}\times{}$31) product-country pairs across a broad representation of
developing and developed markets. In the context of international
diffusion studies, the scale and scope of our data provide a
substantial empirical basis for investigation. For instance, \citet
{Chandrasekaran2007} note that a substantial data basis in this context
should have a sample size of more than 10 countries or 10 products. It
is also important to recall here that the overall time period covered
by our diffusion data spans 25 years from 1980 to 2004 that saw several
interesting and relevant world events in the context of investigating
change in new product diffusion speed over time.

\begin{table}
\caption{Countries in our sample}\label{tablecountries}
\begin{tabular*}{\textwidth}{@{\extracolsep{\fill}}ld{2.2}d{2.2}lld{2.2}d{2.2}@{}}
\hline
\multicolumn{1}{@{}l}{\textbf{Country}}&\multicolumn{1}{c}{\textbf{{\%} Pop.}}&\multicolumn{1}{c}{\textbf{{\%} GNI}}&
&\multicolumn{1}{l}{\textbf{Country}}&\multicolumn{1}{c}{\textbf{{\%} Pop.}}&\multicolumn{1}{c@{}}{\textbf{{\%} GNI}}\\
\hline
Argentina&0.6&0.96&&Italy&0.91&3.06 \\
Australia&0.31&1.19&&Malaysia&0.39&0.47 \\
Austria&0.13&0.49&&Mexico&1.6&1.96 \\
Belgium&0.16&0.62&&Netherlands&0.25&1.01 \\
Brazil&2.88&2.75&&Norway&0.07&0.34 \\
Canada&0.5&1.9&&Philippines&1.29&0.83 \\
Chile&0.25&0.32&&Portugal&0.16&0.39 \\
China&20.19&15.87&&Singapore&0.07&0.22 \\
Denmark&0.08&0.33&&South Korea&0.75&1.91 \\
Finland&0.08&0.3&&Spain&0.67&2.05 \\
France&0.94&3.5&&Sweden&0.14&0.53 \\
Germany&1.28&4.44&&Switzerland&0.12&0.52 \\
Greece&0.17&0.46&&Thailand&0.99&0.98 \\
Hong Kong&0.11&0.43&&United Kingdom&0.93&3.65 \\
India&16.94&6.73&&United States&4.59&22.3 \\
Ireland&0.06&0.25&&{TOTAL}&\multicolumn{1}{c}{57.62}&\multicolumn{1}{c}{80.76}\\
\hline
\multicolumn{7}{l}{Source: \citet{Worldbank2010}}\\
\end{tabular*}
\end{table}

\begin{figure}
\centering
\begin{tabular}{@{}c@{\hspace*{2pt}}c@{}}

\includegraphics{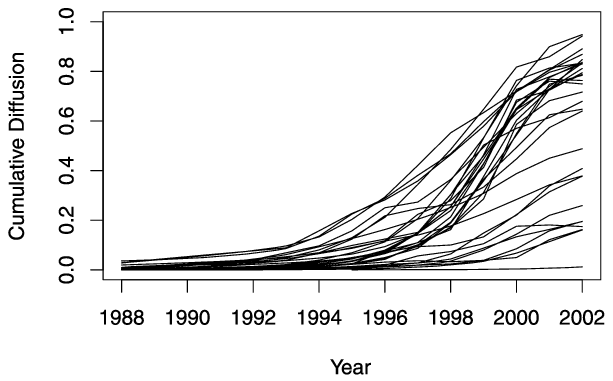}
 & \includegraphics{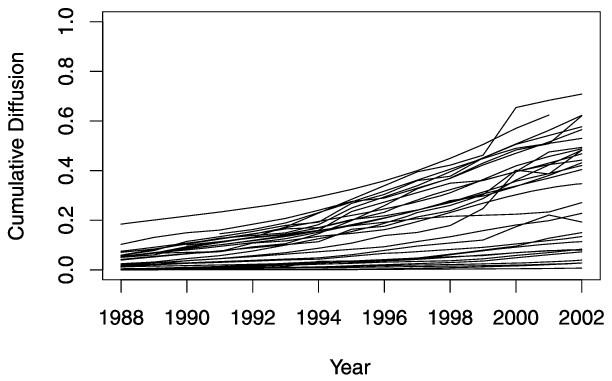}\\
\footnotesize{(a)} & \footnotesize{(b)}
\end{tabular}
\caption{Diffusion trajectories for home computers and cellular phones.
\textup{(a)} Cellular phones. \textup{(b)}~Home computers.}
\label{figEDAsepprod}
\end{figure}

Since our data consists of a wide array of disparate country-product
pairs over a 25-year period, our diffusion data is particularly
interesting, as it contains large variations across countries, across
products and over time. To exemplify such variations, Figure \ref
{figEDAsepprod} plots diffusion trajectories for two of our products
for each of the 31 countries over a common period of 1988--2002. As
evident from the figure, a~comparison across products shows that while
the diffusion of cellular phones was slow to take off, it accelerated
rapidly after the early 1990s. In contrast, the diffusion of home
computers started earlier but its growth has been more gradual. Also,
for a given product, the variation in diffusion patterns across
countries is readily apparent from the figure, with some countries
showing much steeper or faster diffusion than others. For instance, in
the case of home computers, our data shows that the United States
reached 20\% penetration in 1989---five years before the next four
countries (Australia, Canada, Norway and Switzerland), although the
computer was introduced in all five countries around the same time. In
contrast, we find that 10 countries (32\% of our sample) did not reach
20\% penetration by 2004.

For our study, we were able to get data on 22 relevant country-specific
covariates across our sample of 31 countries and for our overall time
period of 1984--2004. Such country-specific covariates are essential to
analyze what drives variation in diffusion speed across countries,
products and over time. Although all our 22 covariates are obviously
time-variant, we were able to find annual data for each of our 31
countries over our entire time window (1980--2004) for only 10 of the
covariates. These covariates are used in our analysis to specifically
capture the temporal variation of diffusion speed within and across
countries. For the other 12 covariates, we were unable to get annual
data for all the countries and every year in our time window. Such
paucity of continuous time-series data on country-specific covariates,
especially in the context of developing countries, is quite typical in
international diffusion studies\vadjust{\goodbreak} [\citet{Chandrasekaran2007}]. As in
other diffusion studies, we use these covariates as time-invariant
country-specific covariates to specifically capture the variation of
diffusion speed across countries [\citet{Talukdar2002}]. The list of the
covariates is given below (the respective years show the particular
year's data used in our analysis for the time-invariant covariates).

\textit{Time-varying covariates}
\begin{itemize}
\item Age dependency ratio: ratio of those in the workforce to those
not in the workforce
\item Consumer price index
\item Electric power consumption (KWH per capita)
\item Gross domestic product (GDP) per capita
\item Household final consumption: average total expenditure per household
\item Internet users (per 1000 people)
\item Labor force participation rate, female: proportion of females in
the labor force
\item Number of telephone mainlines (per 1000 people)
\item Unemployment (\%)
\item Urban population: percent of population living in an urban area
\end{itemize}
\textit{Time-invariant covariates}
\begin{itemize}
\item Daily newspapers: number of newspapers delivered each day, on
average, in 2000
\item Ease of doing business index: how conducive is the regulatory
environment to business, in 2000
\item GINI index: a measure of the inequality of wealth in 2000
\item Households with television: percentage of households with a
television in 1995
\item Individualism index: measure of the degree to which individuals
are integrated into groups [\citet{Hofstede2001}]
\item International migrant stock: number of migrants in the country in 2000
\item International tourism: total tourist entering the country in 1998
\item International voice traffic: minutes of international telephone
calls in 2000
\item Population growth rate in 2000
\item Price basket for residential fixed line: average cost of a
residential fixed telephone line in 2000
\item Pump price for gasoline in 1995
\item Uncertainty avoidance index: deals with tolerance for uncertainty
and ambiguity [\citet{Hofstede2001}]
\end{itemize}
Our study data comes from several international organizations such as
the International Monetary Fund (IMF), International Telecommunications
Union (ITU), the United Nations (UN), the World Bank and the World
Tourism Organization (WTO). Specifically, product adoption and sales
data for each country are based on annual household and respective
industry surveys conducted by various national government agencies. We
obtained the data from the country-level databases of the World Bank,
ITU and from publications by Euromonitor (European and International
Marketing Data and Statistics, various years). As for our various
country specific covariates, the socio-economic development indicator
databases at the UN, WTO and World Bank served as the sources. Our
access to the data is based on specific permission obtained from the
various organizations, so unfortunately we are unable to post the data
as a supplement to this article. Interested parties can contact the
individual organizations for access details.

\section{Methodology}
\label{secMethod}
\subsection{Model}\label{sec3.1}
A review of new product diffusion literature shows that researchers
have essentially used two distinct types of models---the logistic
diffusion model and the Bass diffusion model [\citet
{Chandrasekaran2007}]. The key difference in the structure of the two
models is that while the logistic diffusion model has a single
parameter to capture consumers' propensity to adopt new products, the
Bass diffusion model has two such parameters. Since the speed of the
diffusion process in either of these two models is expressed in terms
of the respective parameters that capture consumers' propensity to
adopt new products, the logistic model---with its single
parameter---provides a more direct and cleaner relationship between its single
parameter for consumers' propensity to adopt a new product and the
speed of the diffusion process [\citet{Fisher1971}]. So, past diffusion
studies focused on diffusion speed, like the one by \citet
{VandenBulte2000} noted earlier, used the logistic diffusion model.
They have also termed the model's single parameter to be the diffusion
growth or speed parameter. Given that the central focus of our study is
diffusion speed, we also use the logistic diffusion model as the base
model for our analysis.\vadjust{\goodbreak}

For the diffusion of a new product in a given country, the basic
logistic diffusion model is given by
%
\begin{equation}
\frac{y(t)}{Y(t-1)}=\lambda\biggl[1-\frac{Y(t-1)}{M(t)\alpha}\biggr] +
\varepsilon(t),
\label{eqLDM}
\end{equation}
where $y(t)$ is the number of adopters in time $t$, $Y(t-1)$ is the
number of cumulative adopters by time $t-1$, $M(t)$ is the population
at time $t$, $\alpha$ is the adoption ceiling parameter (proportion of
the population which will eventually adopt the product), $\lambda$ is
the speed parameter (the main focus of our study), and $\varepsilon(t)$ is
the error term, $\varepsilon(t)\sim N(0,\sigma^2)$. To analyze the
diffusion of 31 electrical household durables in the United States,
\citet{VandenBulte2000} modified the above single product, single
country basic logistic diffusion model into a multi-product, single
country model. Specifically, his model for product $n$ is
\begin{eqnarray}
\frac{y_n(t)}{Y_n(t-1)} &=& \lambda_n\biggl[1 - \frac{Y_n(t-1)}{M(t)\alpha
_n}\biggr] + \sum_{k\in K_{\mathrm{TV}}}\psi_{k}X_{kn}(t) + \varepsilon_n(t),\label
{eqVdB1}\\
\lambda_n &=& \lambda_0 + \sum_{k\in K_{\mathrm{TIV}}}\beta_kX_{kn} + \varepsilon
_{n},\label{eqVdB2}
\end{eqnarray}
where $K_{\mathrm{TV}}$ is the set of time-varying covariates $X_{kn}(t)$, and
$K_{\mathrm{TIV}}$ is the set of time-invariant covariates $X_{kn}$. We augment
the model in equations \eqref{eqVdB1} and \eqref{eqVdB2} in three main
ways. First and most importantly, we allow the speed parameter ($\lambda
$) to vary over the diffusion life cycle of the product. Second, we
modify the model to allow for multiple products and multiple countries.
Additionally, we perform a variable selection procedure to determine
the significant covariates.
\subsection{Multiple products and countries}\label{sec3.2}
To account for the expanded scope of our data in terms of multiple
countries and multiple products, we rewrite the model for country $i$
and product $n$. Because there are only country-specific covariates, we
include a product-specific random effect term, $\tau_{n}$, to account
for variation in speed across products:
\begin{eqnarray}
\frac{y_{in}(t)}{Y_{in}(t-1)} &=& \lambda_{in}\biggl[1 - \frac
{Y_{in}(t-1)}{M_i(t)\alpha_{in}}\biggr] + \sum_{k\in K_{\mathrm{TV}}}\psi
_{k}X_{ki}(t) + \varepsilon_{in}(t),\label{eqstep2-1}\\
\lambda_{in} &=& \lambda_0 + \sum_{k\in K_{\mathrm{TIV}}}\beta_kX_{ki} + \tau_{n}
+ \varepsilon_{in}.\label{eqstep2-2}
\end{eqnarray}
%
\subsection{Time effect}\label{sec3.3}
As is apparent from its specification in equation (\ref{eqVdB2}), the
diffusion model used in the study by \citet{VandenBulte2000} assumes
that the speed parameter for a given product remains constant
throughout its diffusion life cycle. In this context, it is pertinent
to note that the study by \citet{VandenBulte2000} focused on
investigating change in diffusion speed across products introduced in
different time periods. So, for that focus, using a model with a
time-invariant speed parameter over a given product's diffusion life
cycle is reasonable. The assumption of time-invariant speed parameters
in a diffusion model also provides two distinct advantages. For one, it
makes it relatively easier to empirically estimate such models [\citet
{Xie1997}]. It also enables easier derivations of closed-form
expressions for the link between the speed parameter and the amount of
time it takes to go from one penetration level to a higher one [Van den
Bulte (\citeyear{VandenBulte2000})].

At the same time, as noted in our introductory discussion, the
assumption of time-invariant diffusion speed parameters imposes the
restrictive premise that consumers' propensity to adopt a new product
remains constant over its diffusion life cycle. This premise is
conceptually at odds with consumers' adoption process in reality, as
consumers' propensity to adopt a new product is likely to vary over its
diffusion life cycle [\citet{Horsky1990}]. Such variation in consumers'
propensity to adopt a new product will be driven by changes in market
environments over time that influence consumers' risk attitude and
perceived risk of adopting a specific new product and/or new products
in general.

Not surprisingly, even though it makes empirical estimation of
diffusion models more difficult, researchers now recognize the need to
relax the aforesaid restrictive assumption of time-invariant diffusion
speed parameters [\citet{VanEverdingen2005}; \citet{Xie1997}]. Accordingly, to
make our model consistent with this reality, we allow the diffusion
speed parameter $\lambda$ to be time-varying. We should note here that
while using the time-variant diffusion speed parameter makes it more
difficult to derive a~closed-form expression for the link between speed
parameter and the amount of time it takes to go from one penetration
level to a higher one, it is still possible under specific conditions
(see \hyperref[appa]{Appendix A} for details).

To allow the diffusion speed parameter $\lambda$ to be time-varying, we
modify our model specification as follows:
\begin{eqnarray}
\frac{y_{in}(t)}{Y_{in}(t-1)} &=& \lambda_{in}(t)\biggl[1 - \frac
{Y_{in}(t-1)}{M_i(t)\alpha_{in}}\biggr] + \varepsilon_{in}(t),\label
{eqstep3-1}\\
\lambda_{in}(t) &=& f(t) + B_i(t) + \tau_{n} + \tau_{in}(t),\label
{eqstep3-2}\\
B_i(t) &=& \sum_{k\in K}\beta_kX_{ki}(t) + \tau_i(t),
\\
 \tau_{in}(t)&\sim& N(0,\theta_H) ,\qquad \tau_{n}\sim
N(0,\theta_A) ,\qquad \tau_{i}(t)\sim N(0,\theta_B). \label{eqhierEnd}
\end{eqnarray}
As equation (\ref{eqstep3-2}) shows, we decompose the speed parameter
into three components: (1) a common baseline time effect in the form of
a nonparametric function, $f(t)$,\vadjust{\goodbreak} which depends only upon time, (2) a
country-specific term, $\sum_{k}\beta_{ki}X_{ki}(t)$, which includes
all the covariates, and (3) a product-specific random effect $\tau_n$.
The country and product effects on the speed parameter are included
through the $B_i(t)$ and $\tau_n$ terms; so $f(t)$ describes common
time-related effects not specific to any one product or country.
Additionally, any omitted covariates whose values are highly correlated
to the time would also be incorporated in this term (e.g.,
contemporaneous global macro-environmental trends; expected
improvements in quality, price and availability as a product matures).
As such, our model specification allows temporal variation in the speed
parameter to be driven by changes in both the country-specific
covariates as well as by an across-country common time effect. The
covariate $X_{ki}(t)$ for the speed parameter $\lambda$ includes both
the time-varying and time-invariant country-specific covariates.
Therefore, $K$ is the union of $K_{\mathrm{TV}}$ and $K_{\mathrm{TIV}}$. We are able to
combine those covariates because we allow our speed parameter to vary
over time. In contrast to the model specifications (equations (\ref
{eqVdB1}) and (\ref{eqVdB2})) in \citet{VandenBulte2000}, this allows us
to directly capture the effects of the time-varying country-specific
covariates on the speed parameter or consumers' propensity to adopt.

By incorporating a Gaussian residual effect $\tau_{in}(t)$, many of the
conditional distributions for the model parameters are now of standard
form, greatly increasing our computational efficiency. Conditional on
$\lambda_{in}(t)$, equation (\ref{eqstep3-2}) is independent of
$Y_{in}(t)$ and can be written as a standard normal--normal conjugate.
This approach of inducing additional random effects has been taken by
\citet{Holmes2003} and \citet{Liechty2009} in different contexts. We
constrained $\theta_{H}$, the variance of $\tau_{in}(t)$, to be small
as suggested in these papers.
Using Bayesian adaptive regression splines [\citet{Dimatteo2001}],
$f(t)$ is approximated by a cubic spline with $k$ knots in locations
$\xi= (\xi_1,\ldots,\xi_k)$, where $a<t_{(1)}<\xi_1\leq\cdots\leq\xi
_k<t_{(n)}<b$. Also, $b_j(t), j\in\{1,\ldots,k+2\}$ is the $j^{th}$
function in a cubic B-spline basis with natural boundary constraints.
Then $f(t) = \sum_{j=1}^{k+2} \omega_jb_j(t)$ for some $\omega_k, k\in\{
1,\ldots,k+2\}$. The prior distributions are defined as [\citet{Kass1995}]
\begin{eqnarray}
p(k) &=& \operatorname{Poi}(2), \label{eqBARSPriorStart}\\
p(\xi) &=& \operatorname{Unif}(a,b),\\
p(\eta|k,\xi) &=& N(0,1).\label{eqBARSPriorEnd}
\end{eqnarray}
The posterior distributions of the $\xi$ and $\eta$ have dimensions
dependent upon~$k$. To estimate the distributions, we use a reversible
jump MCMC sampler [\citet{Green1995}; \citet{Denison1998}; \citet{Denison2002}]. For each
iteration of the sampler one of three moves are proposed: birth (add a
new knot), death (remove an existing knot), or relocation (move an
existing knot to a new location). This method performs well in our
case, because the smoothness of the function is chosen automatically
and not constrained\vadjust{\goodbreak} to be constant across the domain. If there is a
sharp change point in our data, this method will discover it. For
further information on the implementation of this method, please see
\citet{Wallstrom2008}.
\subsection{Determining the significant covariates}\label{sec3.4}
In the interest of parsimony, we determine which covariates
significantly contribute to the model. The parameter $\gamma_k$ is a
binary variable determining if $\beta_k$ is significantly different
from zero [\citet{George1993}; \citet{George1997}; \citet{Kuo1998}].

The prior distributions for $\gamma$, $\theta_B$ and $\beta$ are
\begin{eqnarray}
p(\gamma_i)&=&\prod_i w_i^{\gamma_i} (1-w_i)^{(1-\gamma_i)},\\
p(\theta_B) &=& \operatorname{IG}(\nu/2,\nu\kappa/2),\\
p(\beta|\theta_B,\gamma) &= &N(0,\theta_B D_\gamma R D_\gamma),
\end{eqnarray}
where $D_\gamma$ is a diagonal matrix and $R$ is a correlation matrix
which we set to be $(X^TX)^{-1}$. The hyperparameters for $\theta_B$
were chosen according to the advice in \citet{George1993}. They
recommend choosing $\kappa= s^2_{LS}$ and then choosing $\nu$ so there
is substantial probability on the interval $(s^2_{LS},s^2_{B_i(t)})$,
where $s^2_{B_i(t)}$ is the sample variance of $B_i(t)$ acquired from a~pilot run. The $i${th} diagonal element of $D_\gamma^2$ is set to
%
\begin{equation}
(D_\gamma^2)_{ii} =
\cases{
0, & \quad $\mbox{when } \gamma_i=0,$\vspace*{2pt}\cr
\upsilon, &\quad $\mbox{when } \gamma_i=1.$
}
\end{equation}
Under those conditions, the marginal distribution of $\beta_i$ is
modeled as
\begin{equation}
p(\beta_i|\theta_B,\gamma) = (1-\gamma_i)I_0 + \gamma_i N(0,\theta
_B\upsilon),
\end{equation}
where $I_0$ is a point mass at 0. Following the suggestions in \citet
{George1997}, we set the value of $\upsilon$ to $\upsilon_\beta/\hat
\theta= 0.122/0.017 = 7.00$, where~$\upsilon_\beta$ is an estimate
consistent with the expected~$\beta$ values (we used the standard
deviation of the least-squares estimates) and $\hat\theta$ is the LS
estimate of $\theta$. In the interest of parsimony, we chose $w$ to be 0.1.

When $\gamma_k$ equals one, the covariate is included in the model.
When it equals zero, the coefficient for that covariate is not
significantly different from zero. Because we draw $\gamma_k$ values
from their posterior distribution in each iteration of the algorithm,
we can determine the posterior probabilities of significance for each
of the covariates by simply finding the proportion of draws which
return a one.

It is possible that the set of selected covariates is dependant upon
the order in which they are sampled [for further exposition, see, e.g.,
\citet{heaton2009}]. To overcome that potential problem, we randomly
determine the order in which the $\gamma_k$ values are sampled in each
iteration and run multiple simultaneous chains to check convergence.\vadjust{\goodbreak}

\subsection{Other prior specifications}\label{sec3.5}
The adoption ceiling ($\alpha$) is bounded both above and below. It is
bounded above by one and below by the maximum cumulative adoption for
the product-country pair observed in our data ($\operatorname{max}(Y_{in}(t))$). The
prior distribution for $\alpha$ is taken to be uniform on that interval.

The precision parameters not involved in the variable selection are
given relatively noninformative prior distributions
\begin{eqnarray} \label{eqvarPrior}
p(\theta_L) &= &\operatorname{Ga}(10^{-5},10^{-5}),\\
p(\theta_A) &=& \operatorname{Ga}(10^{-5},10^{-5}).
\end{eqnarray}
The details of the sampling algorithm are available in \hyperref[appb]{Appendix B}. The
algorithm was implemented in R and the code is available as a
supplement to this article [\citet{Hartman2011a}].

\begin{figure}

\includegraphics{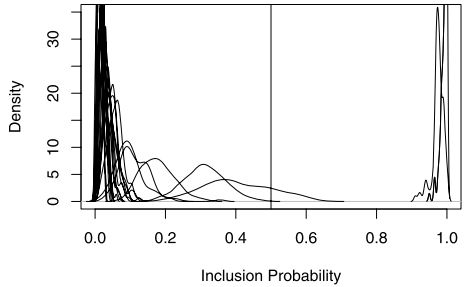}

\caption{Posterior inclusion probabilities for each covariate.}
\label{figVSByCov}
\end{figure}

\section{Results}\label{sec4}
\subsection{Variable selection results}\label{sec4.1}
After running the model with the chosen hyperparameters 100 times, we
obtained the following results. Figure~\ref{figVSByCov} plots kernel
density estimates for the posterior inclusion probabilities for all the
possible covariates. We see that the probabilities are relatively
consistent across runs. The two covariates with all of their mass above
0.5 are the internet penetration level and the consumer price index
(CPI). Electric power consumption only had 19\% of its mass above 0.5
and households with television had less than 1\% above 0.5; so we
conclude that they do not have significant effects. The $\beta$
estimates were consistent across sampler runs, with regular and
unimodal posterior densities. The estimates for the coefficients for
CPI and internet penetration level are $-0.081$ and 0.123 respectively.
Because all the covariates are standardized to have a mean of zero and
a standard deviation of one, the absolute size of the estimates are
less informative than the sign of the estimates.

The variable selection results are consistent with the expected
negative role of the CPI. As the cost\vadjust{\goodbreak} of living rises with inflation
and CPI, it adversely affects consumers' discretionary income and thus
their willingness and ability to pay for new products introduced in the
marketplace [\citet{Horsky1990}; \citet{Talukdar2002}]. The results also follow
the expected positive role of internet access on the speed of a
behavior process, that is, fundamentally driven by information flow among
the adopting population. In this context, it is relevant to point out
that past studies using data from the pre-internet period have included
TV and newspaper penetration levels as covariates of diffusion speed
parameters to recognize the role of mass media on diffusion process
[\citet{Putsis1997}; \citet{Talukdar2002}]. Consistent with those past studies,
our findings underscore the strong role of the new mass medium
represented by the internet, which has fundamentally altered how
consumers and firms search for, store and transmit product related
information, as well as buy and sell products [\citet{Ratchford2007}].
The internet also helps speed up the adoption process by acting as a
product complement for one of the products (home computers) in our study.
\subsection{Prior sensitivity}\label{sec4.2}
\begin{figure}[b]

\includegraphics{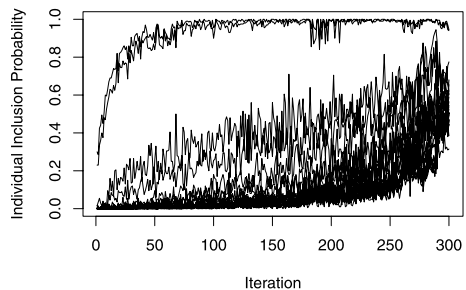}

\caption{Prior sensitivity of variable selection.}
\label{figVSByIter}
\end{figure}

The variable selection results can be highly sensitive to the prior
specification. To test the prior sensitivity of the results, we
performed the analysis ten times for all possible combinations of the
following values for the hyperparameters (300 total runs):
\begin{eqnarray*}
\upsilon&\in&\{1, 5, 7, 10, 15, 20, 25, 50, 100, 500\},\\
w&\in&\{0.1, 0.3, 0.5\}.
\end{eqnarray*}
The inclusion probabilities for all the covariates are plotted in
Figure \ref{figVSByIter}, with the iterations then sorted by the
average inclusion probability over all the covariates.
%

Two of the covariates (internet penetration level and the consumer
price index) are significantly above the others regardless of the
hyperparameter settings. The prior specification obviously has a large
effect on the inclusion probabilities. Table~\ref{tabVSHPMar}\vadjust{\goodbreak} contains
the average inclusion probabilities for various hyperparameter
settings. The hyperparameter $\upsilon$ is negatively related to
inclusion probability and $w$ is positively related, but the chosen
covariates are largely invariant to the settings.\vspace*{-3pt}
%

\begin{table}
\caption{Marginal inclusion probabilities}
\label{tabVSHPMar}
\begin{tabular*}{250pt}{@{\extracolsep{\fill}}lcccc@{}}
\hline
&\multicolumn{3}{c@{}}{$\bolds{w}$}&\\[-6pt]
&\multicolumn{3}{c@{}}{\hrulefill}&\\
$\bolds{\upsilon}$&\textbf{0.1}&\textbf{0.3}&\textbf{0.5}&\textbf{Marginal}\\
\hline
\phantom{00}1&0.20&0.38&0.55&0.37\\
\phantom{00}5&0.16&0.28&0.46&0.30\\
\phantom{00}7&0.15&0.25&0.39&0.26\\
\phantom{0}10&0.13&0.20&0.33&0.22\\
\phantom{0}15&0.13&0.18&0.26&0.19\\
\phantom{0}20&0.11&0.16&0.22&0.17\\
\phantom{0}25&0.11&0.15&0.21&0.16\\
\phantom{0}50&0.09&0.13&0.16&0.13\\
100&0.08&0.11&0.14&0.11\\
500&0.04&0.07&0.09&0.07\\ [3pt]
Marginal&0.12&0.19&0.28&0.20\\
\hline
\end{tabular*}\vspace*{-3pt}
\end{table}
%
\begin{table}[b]
\vspace*{-3pt}
\caption{Adoption ceiling parameter estimates}
\label{tabAdopt}
\begin{tabular*}{250pt}{@{\extracolsep{\fill}}lcc@{}}
\hline
\textbf{Product}&\textbf{Mean}&\textbf{95\% Credible interval}\\
\hline
Cell Phone&0.8001&(0.6069, 0.9840)\\
Home Computer&0.6802&(0.6010, 0.9468)\\
Camcorder&0.7998&(0.6100, 0.9899)\\
CD Player&0.7973&(0.6096, 0.9897)\\
\hline
\end{tabular*}
\end{table}

%
\subsection{Adoption ceiling}\label{sec4.3}
As noted in \citet{VandenBulte2000}, ceiling and speed parameters tend
to be negatively correlated and data with a shorter time series tend to
have lower estimates of the adoption ceiling parameter. However, this
observation is based on diffusion model specifications which impose a
time-invariant structure on the speed parameter. An interesting issue
is whether the observation still holds for a model, as in our study,
which allows the speed parameter to in fact vary over time. In fact,
contrary to the observation, we find a slightly positive correlation
coefficient ($r=0.203$) between the adoption ceiling and speed
parameter estimates for our four products. We also checked whether our
adoption ceiling parameter estimates are in line with those in past
studies, especially for the CD players and camcorders, as those series
have only 7 years of data for each country. The mean and 95\% credible
intervals for the adoption ceiling parameter are in Table \ref
{tabAdopt}.\vadjust{\goodbreak} We find the estimates to be quite consistent with the
findings from other studies [\citet{Talukdar2002}]. In this context, it
is pertinent to point out that the study by \citet{VandenBulte2000},
like our study, also does not find any systematic bias in its estimates
of adoption ceiling parameters for products like camcorders and CD
players with shorter data series.
%

\begin{table}[b]
\caption{DIC results}\label{tableDICtv}
\begin{tabular*}{250pt}{@{\extracolsep{\fill}}lccc@{}}
\hline
\textbf{Time measure}&$\bolds{\bar D}$&$\bolds{P_D}$&\textbf{DIC}\\
\hline
Year since introduction & $-5654.05$ & $1054.10$ & $-4599.95$\\
Calendar year & $-5645.56$ & $1057.42$ & $-4588.15$\\
N.A. (Time-invariant) & $-5638.83$ & $1057.17$ & $-4581.66$\\
\hline
\end{tabular*}
\end{table}
%
%
\subsection{Time component}\label{sec4.4}
Our focal interest in this study is the temporal trajectory of the
diffusion speed parameter. In the context of new product diffusion,
there are two distinct ways we could measure time: calendar year and
year since new product introduction. Also, as noted earlier, our model
specification allows temporal variation in the speed parameter to be
driven by changes in the country-specific covariates as well as a
common time effect captured through the nonparametric function $f(t)$.
For the purpose of testing alternative models within our overall model
structure, we can use either measure of time or remove $f(t)$ from the
model completely. We compared the various alternative models by keeping
the prior settings common and then using DIC [\citet
{Spiegelhalter2002}]. Table \ref{tableDICtv} describes the results of
the model comparison. $\bar D$ is a measure of how well the model fits
the data. $P_D$ is the effective number of parameters which is used as
a complexity penalty. $P_D$ is different from the nominal number of
parameters, especially in hierarchical models. Two models may have the
same number of nominal parameters, but if one model is more
identifiable and precise, it will have a~smaller number of effective
parameters [\citet{Congdon2006}]. DIC is the sum of those two values. In
all cases, a smaller number is better.

Based on the DIC values, we find that a time-varying speed parameter
with time measured in terms of either the calendar year or the year
since product introduction provides a better fit than using a
time-invariant speed parameter. Additionally, using year since
introduction provides the best fit and fewer effective parameters. Our
findings show that modeling the diffusion speed with a time-invariant
speed parameter adversely affects the precision and identification of
the parameters in the model. Even though the nominal number of
parameters is greater when using the number of years since the
introduction, the effective number of parameters in fact gets smaller.
Our findings thus underscore the value for new product diffusion models
in relaxing the typical restrictive assumption of time-invariant\vadjust{\goodbreak}
diffusion speed parameters, and corroborate similar conclusions from
past studies [e.g., \citet{VanEverdingen2005}; \citet{Xie1997}].

\begin{figure}

\includegraphics{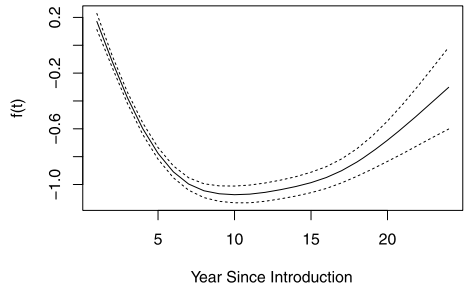}

\caption{$f(t)$ against year since introduction.}
\label{figftPlot}
\end{figure}

Figure \ref{figftPlot} plots the estimated posterior distribution of
$f(t)$ against the number of years since the introduction of the
product in each country. The solid line is the pointwise posterior
mean, and the dashed lines are the 95\% pointwise credible interval
bounds. The plot sheds interesting insights into the patterns of the
diffusion speed parameter $\lambda(t)$ based on the common time effect
induced by the time-correlated product and general macro-environmental
trends. The plot clearly shows that there is a systematic temporal
trend in the speed parameter---thus, in the underlying consumers'
propensity to adopt a new product--over a product's diffusion life
cycle. Specifically, the diffusion speed parameter is found to exhibit
a U-shaped pattern with respect to the time since a new product's
introduction in a country.

Our finding of the U-shaped temporal pattern in the diffusion speed
parameter since a new product's introduction in a country indicates
that consumers' propensity to adopt a new product goes through a
relative drop in its value from the initial phase of the diffusion
cycle before climbing back. While our analysis does not provide any
direct causal insight as to why we see such a temporal pattern in
consumers' propensity to adopt a new product since its introduction in
a country, the pattern appears to be consistent with expectations based
on conceptual notions and empirical evidence in the diffusion
literature. For instance, the initial phase of a new product's
diffusion in a country is primarily driven by the so-called early
adopters or consumer innovators [\citet{Bass1969}; \citet{Chandrasekaran2007}].
The early adopters as a consumer segment represent a relatively small
proportion of the eventual adopters for the new product, but by nature
they have a higher propensity to adopt and are the first consumer
groups to adopt a~new product. Our finding of consumers' propensity to
adopt starting high and then declining is also consistent with the
likely effect of promotion by businesses which accompanies the launch
of a new product in a country [\citet{Golder1997}].\vadjust{\goodbreak} Such promotion
usually has the highest intensity at introduction to generate consumer
awareness and interest for the product, but then declines to a lower
but steady level as businesses rely more on word-of-mouth from the
early adopters. However, the early adopters are followed by the
laggards or late adopters [\citet{Chandrasekaran2007}] with lower
propensity to adopt the new product.

At the same time, as time passes since the introduction, the risk
perception among consumers toward adopting a new product declines with
better quality, price and availability on the supply side. That in turn
will have a~positive impact on the value proposition of the new product
on the demand side, thereby increasing the propensity to adopt the new
product among late adopters [\citet{Horsky1990}]. In our study, these
later years specifically include cellular phones and home computers,
and reflect the time period from the early 1990s to 2004. Both the
products over this time period saw steep decline in price even as their
quality and the scope of their use in everyday life improved
significantly [\citet{Blinder2000}; \citet{Chwelos2008};
\citet{Lawal2002}; \citet{Merkle1998}; \citet{Prensky2001}]. Globally, that has not only made consumers
more appreciative of the value of these products in their everyday life
but also more willing to pay for them [\citet{Talukdar2002}].

As noted earlier, the function $f(t)$ in our model specification of the
diffusion speed parameter will reflect not only the effect of
product-specific covariates that are highly correlated with time but
also the effect of contemporaneous global macro-environmental trends.
In that context, it is relevant and interesting to observe here that
the early time periods in the diffusion cycle of our products span the
early 1980s and early 1990s. This time period saw high levels of
economic anxieties and unemployment across the globe driven by two
recessions and a stock-market crash (1987) in the United States. On the
other hand, the later time periods in the diffusion cycle of our
products span the late 1990s and early 2000s. That time period, in
contrast, witnessed some singular global macro-environmental trends.
For instance, it saw unprecedented trends in economic policy
liberalization and digitization of key aspects of market economies all
over the world [\citet{Gilpin2001}]. These trends had a profound impact
on the global flow of goods, capital and labor---essentially on factors
creating the flat world [\citet{Friedman2005}]. They also continue to
have significant impact on how product information is disseminated and
products are sold by firms as well as how they are searched for and
purchased by consumers. Based on economic rationale [\citet
{Horsky1990}], all the above global macro-environmental trends are
likely to boost consumers' likelihood of adoption of new products in
general, and especially of cellular phones and home computers---consistent
with our findings discussed earlier about the temporal
pattern of $f(t)$ in the later stages of the diffusion cycle in Figure~\ref{figftPlot}.

\begin{figure}
\centering
\begin{tabular}{@{}c@{\hspace*{2pt}}c@{}}

\includegraphics{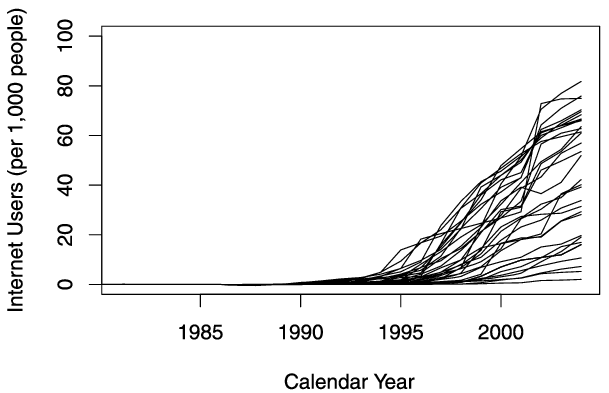}
 & \includegraphics{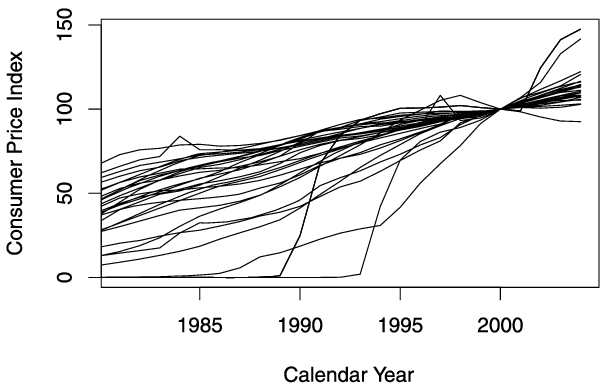}\\
\footnotesize{(a)} & \footnotesize{(b)}
\end{tabular}
\caption{Variation in selected covariates across countries and over time.
\textup{(a)} Internet users (per 1000 people). \textup{(b)} Consumer price index.}
\label{figbetaTV}
\end{figure}

Further, in our model specification, the temporal variation in the
diffusion speed parameter is not just\vadjust{\goodbreak} driven by the common time effect
captured through the function $f(t)$. It is also driven by changes in
the country-specific time-varying covariates. Since the two covariates
(internet penetration level and consumer price index) identified
through our variable selection analysis are both time-varying, we thus
need to include them when looking at the time trend patterns of the
diffusion speed parameter. Figure \ref{figbetaTV} plots the two
selected covariates against calendar year for each country. Consumer
price index (CPI) is calibrated by setting the year 2000 value to 100.
As evident from Figure \ref{figbetaTV}, CPI has increased over our
time window. As for internet penetration level, it first grew above
zero in 1989, but did not dramatically increase until the introduction
of Netscape in 1995 [\citet{Friedman2005}].
%

\begin{figure}[b]

\includegraphics{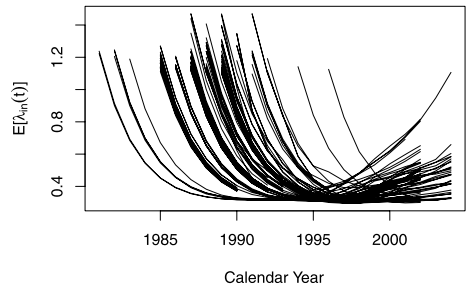}

\caption{Expected trajectory of the diffusion speed parameter.}
\label{figTVfull}
\end{figure}

\begin{figure}
\centering
\begin{tabular}{@{}c@{\hspace*{2pt}}c@{}}

\includegraphics{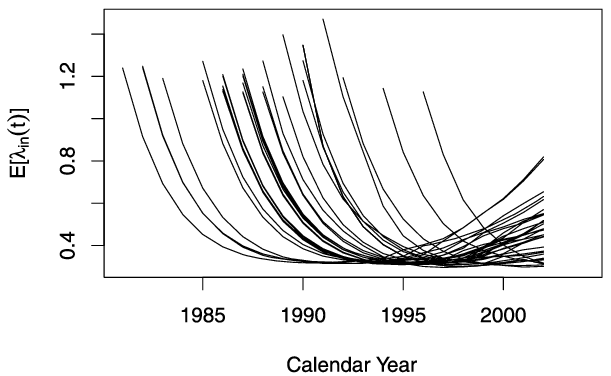}
 & \includegraphics{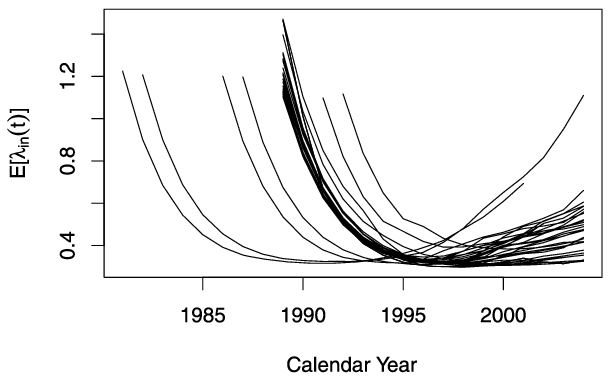}\\[3pt]
\footnotesize{(a)} & \footnotesize{(b)}\\

\includegraphics{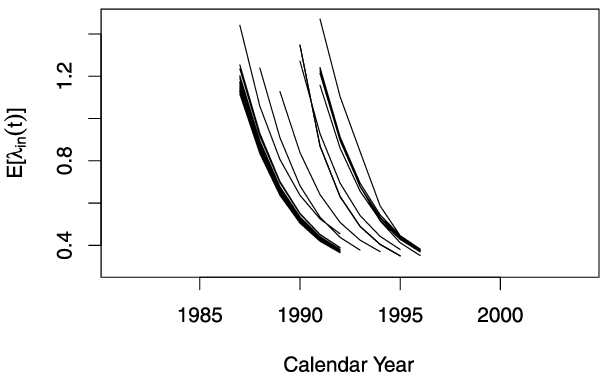}
 & \includegraphics{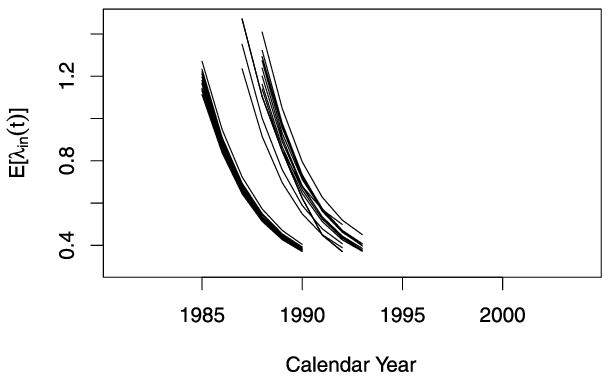}\\
\footnotesize{(c)} & \footnotesize{(d)}
\end{tabular}
\caption{Expected trajectory of the diffusion speed parameters for each
product. \textup{(a)} Cell phones. \textup{(b)} Home computers. \textup{(c)} Camcorders. \textup{(d)} CD players.}
\label{figIndivProdSpeed}
\end{figure}

Now that we have all the individual time-varying components, we can
exponentiate the sum of the components to examine how the diffusion
speed parameter, $\lambda_{in}(t)$, has changed over calendar time.
Figure \ref{figTVfull} plots the expected value of $\lambda_{in}(t)$
over the calendar time covered by our analysis across the four
products. Since the time period covered for each product is different,
it is also instructive to look at similar plots (see Figure \ref
{figIndivProdSpeed}) separately for each product.\vadjust{\goodbreak} Looking at the plots
in Figures \ref{figTVfull} and \ref{figIndivProdSpeed}, it is
apparent that there are two separate time periods corresponding to two
distinct time-trends in the expected value of the speed parameter. From
1980 to the early 1990s, the countries had significant variations in
terms of CPI but very little in terms of internet penetration levels.
At the same time, we find the expected value of the speed parameter for
each product and country pair during this time period to be relatively
parallel. That suggests that during that period the expected value is
dominated by the $f(t)$ term. Consumers' propensity to adopt new
products during this time period declined and was primarily driven by a
common time effect across the countries rather than by any
country-specific covariate effects.

As for the time period between the mid-1990s to 2004, our analysis
covers the later stages of the diffusion cycle for two of the products,
viz., cell phones and home computers. Except for a few exceptions that
we note below, we find the expected value of the speed parameter for
each country for both these products not only reversing direction but
also showing distinctive differences in that upward trend. Coupled with
our earlier finding in Figure \ref{figftPlot}, this finding suggests
that the positive impacts of common time effect and country-specific
internet penetration effects dominated the negative impacts of
country-specific CPI effects on consumers' propensity to adopt new
products\vadjust{\goodbreak} from the mid-1990s to 2004. We should note here that there are
a~few exceptions to the observed U-shaped temporal pattern in the
diffusion speed parameter over the entire time period from the 1980s to
the 2000s. Specifically, the expected speed parameters for five home
computer (Argentina, Brazil, Greece, India and the Philippines) and
four cellular phone (Argentina, Brazil, India and the Philippines)
series start high and drop to a low, but do not significantly increase
toward the end of our data. Interestingly, both Argentina and Brazil
had the highest inflation of the countries in our set. Such high
inflation is expected to depress the speed parameter through its
negative relationship with CPI. The other three countries also had high
inflation (all in the top seven of our set), but they were mainly
affected by a late introduction year (they were late in home computers
and actually were the last three to introduce cell phones).

\begin{table}
\caption{Expected value of the diffusion speed parameter for home computers}
\label{tablespeed2004}
\begin{tabular*}{\textwidth}{@{\extracolsep{\fill}}lclc@{}}
\hline
\textbf{Country}&\textbf{Expected value}&\textbf{Country}&\textbf{Expected value}\\
\hline
Argentina&0.482&Italy&0.474\\
Australia&0.494&Malaysia&0.473\\
Austria&0.474&Mexico&0.534\\
Belgium&0.466&Netherlands&0.491\\
Brazil&0.559&Norway&0.491\\
Canada&0.487&Philippines&0.510\\
Chile&0.509&Portugal&0.493\\
China&0.478&Singapore&0.493\\
Denmark&0.485&South Korea&0.504\\
Finland&0.508&Spain&0.468\\
France&0.457&Sweden&0.512\\
Germany&0.467&Switzerland&0.480\\
Greece&0.546&Thailand&0.464\\
Hong Kong&0.495&United Kingdom&0.481\\
India&0.588&United States&0.502\\
Ireland&0.461&&\\
\hline
\end{tabular*}
\end{table}

For an illustration of the estimated values of the diffusion speed
parameter across the countries, we show in Table \ref{tablespeed2004}
the expected values of the speed parameter for home computers by
country. All of the expected values of the speed parameter are between
0.35 and 0.60, with an average value of 0.49 and a standard deviation
of 0.03. For our entire data set, the expected values of the speed
parameter have a mean of 0.55 and a standard deviation of 0.27 across
all the 124 product-country pairs. These mean estimates of the
time-varying speed parameter are well within the range seen in the past
studies [\citet{Sultan1990}; \citet{VandenBulte2004}]. It is also interesting
and pertinent to note here that while there are very few past studies
of diffusion in developing countries, those studies found that\vadjust{\goodbreak}
developing countries often show higher speed at comparable stages of
their diffusion cycle [\citet{Talukdar2002}; \citet{Takada1991}]. The accepted
rationale is that developing countries generally experience a lagged
national introduction of a new product. Such lag in fact has a positive
effect on diffusion speed, as it means that some developing countries
had not only conducive macroeconomic conditions for adoption but also
the advantage of less adoption risk perception (through better product
price and/or quality) by their consumers at comparable stages of the
diffusion cycle [\citet{Chandrasekaran2007}; \citet{Takada1991}]. As Table \ref
{tablespeed2004} shows, we also find several developing countries like
India and Brazil exhibiting relatively higher values of the diffusion
speed parameter for home computers.

\section{Conclusion}\label{sec5}
Understanding the dynamic nature of new product diffusion speed is
essential for global marketing managers to make informed decisions. Our
paper provides one of the most comprehensive studies of international
new product diffusion speed from both a substantive and methodological
perspective.
First, recent reviews of the new product diffusion literature
underscore the need for studies that expand the scope to include
developing countries [\citet{Chandrasekaran2007}; \citet{Peres2010}]. Our study
works to fill that need by using a data set that includes 31 developed
and developing countries that account for about 80\% of the global
economic output and 60\% of the global population. The time period
(1980--2004) analyzed includes several global events---for example, the
popular emergence of the internet---that are relevant in the context
of investigating change in international new product diffusion speed
over time. Second, our study uses a novel methodology to analyze the
changes in diffusion speed. Specifically, we use a semiparametric model
to allow the diffusion speed parameter to be time-variant. We also use
a variable selection procedure to develop a parsimonious model from the
multitude of potential covariates available in an international
diffusion study.

Taken together, the scope of our data and our methodology enables us to
shed insights into several important issues that are hitherto missing
from the extant literature on new product diffusions. By relaxing the
assumption of a time-invariant speed parameter over the diffusion cycle
of a new product [\citet{VandenBulte2000}], we show that the speed
parameter is generally higher at its introduction, falling to a low in
the middle of the diffusion process, and increasing again in the later
stages. Also, our global data set allows us to show that this
phenomenon occurs not only in developed nations but also in developing
ones. Putting our findings in a broader context, we find that the
global new product diffusion speed increased from the mid-1990s to
2004, a~time period which saw sustained global economic expansion
driven by a~high level of globalization and the ushering in of the
digital age [\citet{Friedman2005}]. Through our variable selection
analysis, we find that the internet penetration level and the consumer
price index in a country are highly associated with the speed of new
product diffusion.

In conclusion, given the scope of our data and our methodology, we have
been able to shed several interesting insights into new product
diffusion speed. We hope our research serves as an impetus for more
work in international new product diffusion. An example of future
research directions from a methodological perspective could be to relax
the assumption of a time-invariant adoption ceiling parameter that has
been used in both past studies and this study. Additionally, the scope
of our data could be expanded to include product-specific covariates.
Although collecting such information in itself---especially for
developing countries---will be quite challenging, the collected data
can be easily incorporated into our hierarchical model structure above
the $\tau_n$ terms. While the function $f(t)$ incorporates
product-specific covariates highly correlated with time, more data
would allow our model to account for those covariates which are
relatively uncorrelated with time.

\begin{appendix}\label{app}

\section{Calculation of the time from one penetration
level to another}\label{appa} 
The speed parameter [denoted in this paper by $\lambda$, and by $\beta$
in \citet{VandenBulte2000}] in the logistic diffusion model
conceptually represents consumers' propensity to adopt a new product
through a social-contagion based diffusion process. The analytical
structure of the standard logistic diffusion model is given by
%
\begin{equation}
x(t) = \lambda F(t-1) [M-X(t-1)], \label{eqA1}
\end{equation}
where $X(t)$ is the cumulative number of adopters at time $t$, $x(t)$
is the incremental adoption at time $t$, $M$ is the number of eventual
adopters, and $F(t) = X(t)/M$ is the penetration level at time $t$. The
speed parameter $\lambda$ affects the slope and displacement of the
logistic diffusion curve, and has thus an intrinsic relationship to the
speed of the underlying diffusion process. To see this relationship,
from equation (\ref{eqA1}) we get
\begin{eqnarray}
\frac{X(t) - X(t-1)}{M} &=& \lambda\frac{X(t-1)}{M}\biggl[1-\frac
{X(t-1)}{M}\biggr],\\
F(t) - F(t-1) &=& \lambda F(t-1)[1-F(t-1)],\\
\frac{dF(t)}{dt} &=& \lambda F(t) [1-F(t)],\\
\int\lambda \,dt &=& \int\frac{dF(t)}{F(t)[1-F(t)]}. \label{eqA2}
\end{eqnarray}
Assuming the speed parameter $\lambda$ is time-invariant, it is quite
easy to solve the integral in equation (\ref{eqA2}) to get a closed-form
solution for the relationship between~$\lambda$ and the speed of the
underlying diffusion process. For instance, the time ($t_2 - t_1$) that
it takes for the diffusion process to go from one penetration level,
$p_1$, to a higher level, $p_2$, is equal to
\begin{eqnarray}
t_2 - t_1 &=& \lambda^{-1}\int_{p_1}^{p_2}\biggl[\frac{1}{F(t)[1-F(t)]}
\biggr] \,dF(t),\\
\Delta t &=& \lambda^{-1} \operatorname{ln}\biggl[ \frac{(1-p_1)p_2}{(1-p_2)p_1}
\biggr]. \label{eqA3}
\end{eqnarray}
On the other hand, assuming that the speed parameter $\lambda$ is
time-variant, the intrinsic mapping of the parameter $\lambda$ to the
speed of the diffusion process is more difficult to derive as a
closed-form solution like equation (\ref{eqA3}), because the solution
comes from equation (\ref{eqA4}) rather than equation (\ref{eqA2}):
%
\begin{equation} \int\lambda(t) \,dt = \int\frac{dF(t)}{F(t)[1-F(t)]}.
\label{eqA4}
\end{equation}
Obviously, equation (\ref{eqA4}) can still be solved numerically.
However, the availability of a closed-form solution will depend on the
specific functional form of $\lambda(t)$. For instance, if $\lambda(t)$
is specified as a linear function of $t$, meaning $\lambda(t) = \lambda
t$, the solution will be
%
\begin{equation} \Delta t = \frac{2}{\lambda(t_1 + t_2)} \operatorname{ln} \biggl[ \frac
{(1-p_1)p_2}{(1-p_2)p-1}\biggr]. \label{eqA5}
\end{equation}
\section{Posterior computation}\label{appb} 
Samples from the posterior distributions of the parameters are drawn
using the following algorithm.
\begin{enumerate}[1.]
\item[1.] Draw the precision parameters from the following full conditional
distributions:
\begin{eqnarray}
s_1^2 &=& \sum^N_{n=1}\sum^I_{i=1}\sum_{t\in T_{in}}\!\biggl\{\frac
{y_{in}(t)}{Y_{in}(t\!-\!1)\lambda_{in}(t)[1\!-\!{Y_{in}(t\!-\!1)}/{(\alpha_{in}M_i(t))}]}\biggr\}^2,
\label{eqprecPostStart}\\
p(\theta_L|\cdot) &=& \operatorname{Ga}\biggl(10^{-5} + \frac{\sum^N_{n=1}\sum
^I_{i=1}T_{in}}{2},10^{-5} +\frac{s_1^2}{2}\biggr),\\
s_2^2 &=& \sum^N_{n=1}\sum^I_{i=1}\sum_{t\in T_{in}}\{[\lambda
_{in}(t)] - f(t) - \tau_n - B_{i}(t)\}^2,\\
p(\theta_A|\cdot) &=& \operatorname{Ga}\biggl(10^{-5} + \frac{N}{2},10^{-5} +\frac{\sum
_{n=1}^N\sum_{t\in T_i}\tau_n^2}{2}\biggr),
\\
\quad p(\theta_B|\cdot) &=& \operatorname{Ga}\biggl(10^{-5}\!+\!\frac{I}{2},
10^{-5}\!+\!\frac{\sum_{i=1}^I\sum_{t\in T_i}(B_{i}(t)\!-\!\sum_{k\in
K}\gamma_k\beta_k X_{ki})^2}{2}\biggr).\hspace*{-30pt}
\end{eqnarray}
\item[2.] Draw the random effects from
\begin{eqnarray}
p(\tau_n|\cdot) &=& N\biggl(\frac{N\theta_H\sum(\lambda_{in}(t)-f(t) -
B_{i}(t))}{\theta_A + N\theta_H},\theta_A + N\theta_H\biggr),\\
p(B_i(t)|\cdot) &=& N(\mu_B,\theta_B + N\theta_H),\\
\mu_B &=& \frac{\theta_B\sum\gamma_kX_k(t)\beta_k + I\theta_H\sum
(\lambda_{in}(t)-f(t) - \tau_n)}{\theta_B + N\theta_H}.
\end{eqnarray}
\item[3.] Draw $\gamma$ from
\begin{eqnarray}
\tilde{Y} &=&
\left[\matrix{ B_i(t)\vspace*{2pt}\cr 0
}\right],\qquad
 \tilde{X}_\gamma=
\left[\matrix{ X\vspace*{2pt}\cr D_\gamma R D_\gamma
}\right],
\\
S^2_\beta&= &\tilde{Y}^T\tilde{Y}-\tilde{Y}^T\tilde{X}_\gamma(\tilde
{X}_\gamma^T\tilde{X}_\gamma)^{-1}\tilde{X}_\gamma^T\tilde{Y},
\\
p(\gamma|\cdot)&=&|\tilde{X}_\gamma^T\tilde{X}_\gamma
|^{-1/2}|D_\gamma R D_\gamma|^{-1/2}
\nonumber
\\[-8pt]
\\[-8pt]
\nonumber
&&{}\times(2\cdot10^{-5}S^2_\beta
)^{-(\sum_{i=1}^I T_i+10^{-5})/2}p(\gamma).
\end{eqnarray}
\item[4.] Draw $\beta$ from
\begin{equation}
p(\beta|\cdot) = N\bigl((X^TX + D_\gamma R D_\gamma)^{-1}X
B_i(t),(X^TX + D_\gamma R D_\gamma)^{-1}\bigr).
\end{equation}
\item[5.] Draw $f(t)$ using the Bayesian adaptive regression splines
algorithm described in \citet{Wallstrom2008}.
\item[6.] Propose a new $\alpha_{in}$ from its prior distribution ($p(\alpha
_{in})\propto1_{[Y_{in}(T_{in}),1]}$) and use a Metropolis--Hastings
step to compute the acceptance probability using the following
likelihood:
%
\begin{equation}
p(Y|\alpha_{in},\cdot) \propto N\biggl[\frac
{y_{in}(t)}{Y_{in}(t-1)} - \lambda_{in}(t)\biggl[1 - \frac
{Y_{in}(t-1)}{M_i(t)\alpha_{in}}\biggr]\Big|0,\theta_L\biggr]. \label{eqACpost}
\end{equation}
\item[7.] Propose a new $\lambda_{in}$ by adding white noise to the
previous value. Use a~Metropolis--Hastings step to calculate the
acceptance probability using the following likelihood:
\begin{eqnarray}
p(\lambda_{in}(t))& \propto& N\biggl[\frac{y_{in}(t)}{Y_{in}(t-1)}
- \lambda_{in}(t)\biggl[1 - \frac{Y_{in}(t-1)}{M_i(t)\alpha_{in}}
\biggr]\Big|0,\theta_L\biggr] \nonumber\\
&&{}\times N[\lambda_{in}(t) - f(t) - B_i(t) - \tau_{n}|0,\theta
_H]\\
&&{}\times \operatorname{Ga}(\lambda_{in}(t)|0.001,1000).\nonumber
\end{eqnarray}
\item[8.] Repeat steps 1--7 until convergence.
\end{enumerate}
\end{appendix}

\section*{Acknowledgments}
The authors are listed in alphabetical order and all contributed
equally. They would like to thank the editors and reviewers for their
effort and advice which vastly improved the paper. They are also
grateful to Craig Meisner (World Bank) and Azucena Pernia (United
Nations World Tourism Organization)\vadjust{\goodbreak} for their help in providing much of
the data used in this study. They also thank the participants at the
2009 Joint Statistical Meeting, the University of Waterloo and the
University of Wisconsin for their valuable comments.

\begin{supplement}[id=suppA]
\stitle{R Code}
\slink[doi]{10.1214/11-AOAS519SUPP} 
\slink[url]{http://lib.stat.cmu.edu/aoas/519/supplement.R}
\sdatatype{.R}
\sdescription{This supplement contains the R code from ``Investigating
International New Product Diffusion Speed: A Semiparametric Approach.''}
\end{supplement}

%

%


\printaddresses

\end{document}